\documentclass[twocolumn,twoside,letterpaper,11pt]{article}
\usepackage{graphicx,xrrc,natbib}
\usepackage{times}
\def\xmm{{\it XMM-Newton\/}}

\def \h50 {h$_{50}$}

\begin{document}

% The title is completely in capital letters and has no ending period

\title{XMM-NEWTON OBSERVATIONS OF THREE HIGH REDSHIFT RADIO GALAXIES}
% We use initials, no first names, for the authors - see example for
% addresses (one line, comma seperated) and how to handle multiple
% authors per institute (comma's and no 'and's before any author):

% The first example is for one or two authors. The second example is
%for 3+ authors. Comment out the one you don't want to use.

%----------------------Begin One or Two Authors-----------------------------
%
%\author{    T. H. E. First Author                        } % author(s)
%\institute{ National Radio Astronomy Observatory         } % the institute
%\address{   P.O. Box O, Socorro, NM 87801, U.S.A.        } % street address
%\email{     xraydio@aoc.nrao.edu                         } % email addresses
%
%
%\author{    T. H. E. Second}
%\institute{ Imaginary Institute}
%\address{   P.O. Box 8000, 9999 ZZ Somewhere}
%         %  *** resuse \address to split in two lines ***
%\address{   The Netherlands}
%\email{     second@imagine.org}
%
%----------------------End One or Two Authors---------------------------

%----------------------Begin Three or More Authors----------------------

\author{    E. Belsole,	D.M. Worrall, M. J. Hardcastle}	 % author(s)
\institute{ Department of Physics - University of Bristol        } % the institute
\address{   Tyndall Avenue, Bristol BS8 1TL, UK        } % street address
\email{     e.belsole@bristol.ac.uk, D.Worrall@bristol.ac.uk, M.Hardcastle@bristol.ac.uk } % email addresses

%\author{    T. H. E. Second, T. H. E. Third, T. H. E. Fourth}
%\email{     second@imagine.org, third@imagine.org, fourth@imagine.org}

%----------------------End Three or More Authors----------------------

\maketitle

\abstract{We present results on the physical states of three high-redshift
powerful radio galaxies (3C\,292 at $z=0.7$, 3C\,184 at $z=1$, and 3C\,322 at
$z=1.7$). They were obtained by combining radio measurements with X-ray 
measurements from \xmm\ that separate spectrally and/or spatially
radio-related and hot-gas X-ray emission. Originally observed as part of a 
programme to trace clusters of galaxies at high redshift, none of the 
sources is found to lie in a rich X-ray-emitting environment similar
to those of some powerful radio galaxies at low redshift, although the
estimated gas pressures are sufficient to confine the radio lobes.
The weak gas emission is a particularly interesting result for 3C\,184, where 
a gravitational arc is seen, suggesting the presence of a very massive cluster.
Here {\em Chandra} data complement the \xmm\ measurements in spatially
separating X-ray extended emission from that associated with the
nucleus and rather small radio source.
3C\,292 is the source for which the X-ray-emitting gas is measured with 
the greatest accuracy, and its temperature of 2 keV and luminosity of 6.5 10$^{43}$ 
erg s$^{-1}$ are both characteristic of a poor cluster.  This source allows the most
accurate measurement of inverse-Compton X-ray emission associated with
the radio lobes.
In all structures where the magnetic-field strength can be estimated through 
combining measurements of radio-synchrotron and inverse-Compton-X-ray emission,
the field strengths are consistent with sources being in a state of minimum 
total energy.}

\section{Introduction}

Powerful radio galaxies are visible to high redshift and hence can be used as tracers of  large-scale structures. It has been hypothesised that powerful radio galaxies may represent a means to the discovery of high-redshift clusters of galaxies (e.g Le F\`evre et al. 1996, Fabian et al. 2001) through their X-ray emission. The new X-ray satellites allow us to test this hypothesis. Several mechanisms are responsible for the X-ray emission from radio galaxies. Thermal emission from a hot atmosphere surrounding the galaxy gives information about the large scale gas distribution and the interaction between the expanding radio source and the confining medium. Unresolved emission from the AGN, which can be thermal or non-thermal,  probes the physics near the central engine. The radio jet and lobes, which also emit in X-rays by the IC and/or synchrotron processes, allow us to put constraints on the energy budget.
In this paper, we present results from \xmm\ observations of three high-redshift radio galaxies. These high-sensitivity data allow us spectrally and spatially to separate different components, and to use the comparison of X-ray and radio data  to constrain the physics of the radio galaxies and their environment. In this work we adopt a cosmology with $H_{\rm 0}$ = 70 km s$^{-1}$ Mpc$^{-1}$, $\Omega_{\rm m}$ = 0.3, $\Omega_{\Lambda}$ = 0.7. If not otherwise stated, errors are quoted at 1$\sigma$ confidence level.

\section{Observations and data preparation}
3C\,184 was observed with \xmm\ for a total of 113 ks between September 2001 and October 2002. After screening the data for high background we have a total useful exposure time of 54 ks for MOS and 16.5 ks for pn (see Belsole et al. 2004 for details).
3C\,292 was observed in October 2002 with the MEDIUM filter for 34 ks. The background screening leaves 20 useful ks for scientific analysis.
The \xmm\ observation of 3C 322 was obtained in May 2002 and was highly contaminated by background flares, which reduce the useful exposure time to 10 ks.

 All radio data used in this analysis are from the NRAO Very Large
Array (VLA), and with
the exception of the low-resolution map of 3C\,292  the maps  were made by us
within {\sc aips} in the standard manner. Observations for 3C\,292 are at 1.4 GHz (Array A) and at 8.5 GHz (Array C). We used 1.4 GHz and 4.87 GHz archive data (Array A) for 3C\,322.

For the three sources we performed a spatial and spectral analysis. We minimised particle contamination by limiting the spectral range for the radial 
profile analysis to 0.2-2.5 keV. This selection criterion should also reduce the 
amount that the wings of the central point source contaminates the extended emission,
if the source displays a hard spectrum. The point-source component was modelled with and analytical function of the Point Spread Function (PSF) for each instrument. Cluster components were modelled with a $\beta$-model convolved with the PSF. Spectra were extracted from vignetting corrected events lists, because this allows us to use the central response file and effective area for each camera. 
Since all three sources are small relative to the EPIC field of view, we took local  background estimates. Moreover, all spectral components used to fit the data were assumed to be absorbed by the Galactic column in the direction of each source.

\section{Results}
\subsection{3C 184}
\begin{figure}[!]
  \begin{center}
    \includegraphics[width=\columnwidth]{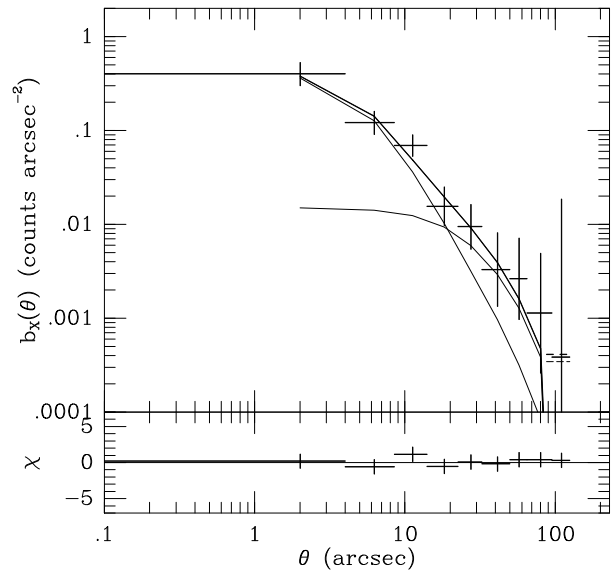}
% Other \includegraphics options include: height, width, angle, scale
    \caption{\small Radial profile obtained from the MOS1 camera in the energy range 0.2 -2.5 keV. The dark grey lines corresponds to the respective contribution of the pointlike model and the $\beta$-model. The best-fitting, combined model is plotted in black on the data points. The lower panel shows the residuals in terms of $\chi$.}
    \label{fig:fig1}
  \end{center}
\end{figure}

The X-ray image displays a rather compact emission, with some indication of extended emission at least out to 40 arcsec. The radio source is very small ($6\times2$ arcsec) and the core of the radio source coincides with the peak of the X-ray emission in the whole energy band.

We extracted a radial profile of the source, which suggests extended emission out to $\sim88$ arcsec (700 kpc). The MOS radial profile is well fitted with a point source plus a $\beta$-model of $\beta$=0.66 and core radius $r_c$= 125 kpc. From the $\beta$-model fit and assuming k$T \sim 3.5$ keV we estimate a bolometric $L_{\rm X} = 6.4\times10^{43}$ erg s$^{-1}$. The  external pressure at the radio lobes (6 arcsec) is $\simeq 3.0\times10^{-12}$ Pa. 

The X-ray spectrum is best fitted with a three-component model. The first component is given by  a soft power law of photon index $\Gamma = 1.5$, representing emission from the lobes and hotspots. The best fit parameters of this non-thermal component were derived using the higher spatial resolution, 20 ks {\em Chandra} observation (see Belsole et al. 2004). With this model, we find a 1 keV X-ray flux density  in the lobes of 0.2 nJy, in agreement with the prediction from SSC and IC scattering of the CMB and iR photons for a source in equipartition.

The second component is represented by an absorbed hard power law with photon index $\Gamma = 1.35$ and N$_{\rm H} = 4.9 \times 10^{23}$ cm$^{-2}$, and finally we include a thermal component of  k$T$ = 3.6 keV.  

\begin{figure}[!]
  \begin{center}
    \includegraphics[width=\columnwidth, angle=0]{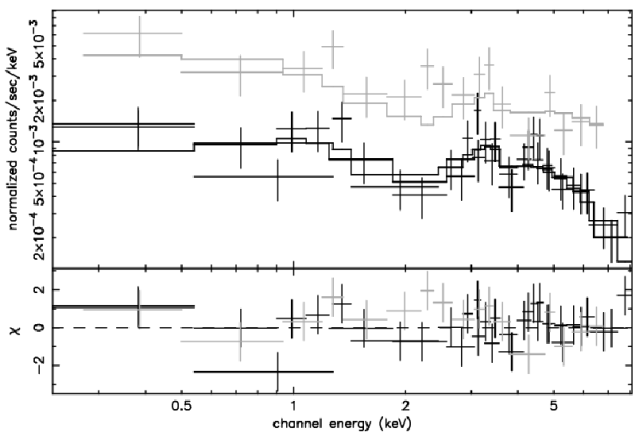}
% Other \includegraphics options include: height, width, angle, scale
    \caption{\small Background subtracted spectrum and folded model of the central 40 arcsec of 3C 184. The pn is in grey, the MOS in black. $\chi$ deviation from the model is in the bottom panel.}
    \label{fig:fig2}
  \end{center}
\end{figure}

We conclude that half of the  X-ray emission of 3C\,184 comes from an absorbed component 
which is associated with the central AGN. Some of the soft emission is radio related and produced by IC scattering. Most of the emission at low energy is thermal and associated with a cluster environment of k$T$ = 3.6 keV and bolometric $L_{\rm X} = 8.4\times10^{43}$ erg s$^{-1}$ (spatial and spectral results agree within the uncertainties).
The cluster external pressure is insufficient to confine the lobes of the radio galaxy which will continue to expand. 

Further support for the existence of  a cluster environment comes from the detection of an arc with the HST (Deltorn et al. 1997), suggesting a virial mass of 7.7 $10^{14} h_{70}^{-1} M_{\odot}$. If we adopt  the proton density calculated from the $\beta$-model, we estimate a gas-mass within a cylinder of radius 5 arcsec (the distance of the arc) integrated along the line of sight of  $\sim1.2\times 10^{11}$ $M_{\odot}$. We compared this value with the total mass calculated in the same cylinder (and scaled for our cosmology) of Deltorn et al. (1997), e.g. $\sim(2.1\pm0.9)10^{13}$ M$_{\odot}$, and found  a gas-mass to total-mass ratio of order 0.01. This is at least  a factor of 10 lower than that observed for clusters of galaxies at lower redshift.  Errors on this estimate are large, and the largest uncertainty is likely to come from the $\beta$-model parameters. The best estimate for the core-radius seems to suggest a  rather shallow potential well with respect to other clusters at the same redshift (e.g. Vikhlinin et al. 2002). Since the spectral fit also implies an under-luminous object, we can speculate that the cluster around 3C\,184 is somehow peculiar in its relatively high temperature but low luminosity (and low gas fraction) if compared to lower redshift galaxy clusters. The fact that the radio source is small and young and expanding might explain the unusually high temperature of the external environment (see also Croston et al. 2004, this meeting).

\subsection{3C 292}
The image shows extended X-ray emission, mainly aligned with the radio lobes, but less extended in the X-ray than in the radio (1.4 GHz). A radial profile fitting (excluding the lobes) gives $\beta$=0.8, $r_c$=19.7 arcsec (140 kpc) and a central intensity of I$_0$=2.3 10$^{-5}$ pn-cnts s$^{-1}$ arcsec$^{-2}$. 
\begin{figure}[!]
  \begin{center}
    \includegraphics[width=\columnwidth, angle=0]{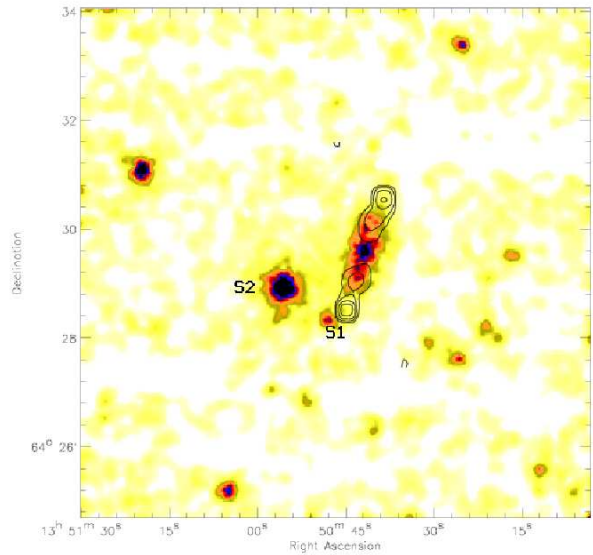}
% Other \includegraphics options include: height, width, angle, scale
    \caption{\small XMM/EPIC adaptive smoothed image of 3C\,292. Contours are from the VLA 1.4 GHz image.}
    \label{fig:fig3}
  \end{center}
\end{figure}

%% Figure:  %%%%%%%%%%%%%%%%%%%%%%%%%%%%%%%%%%%%%%%%%%%%%%%%%%%%%%%%%%%%%%%%%%%%%%%
\begin{figure}
  \begin{center}
   \includegraphics[scale=0.35,angle=-90]{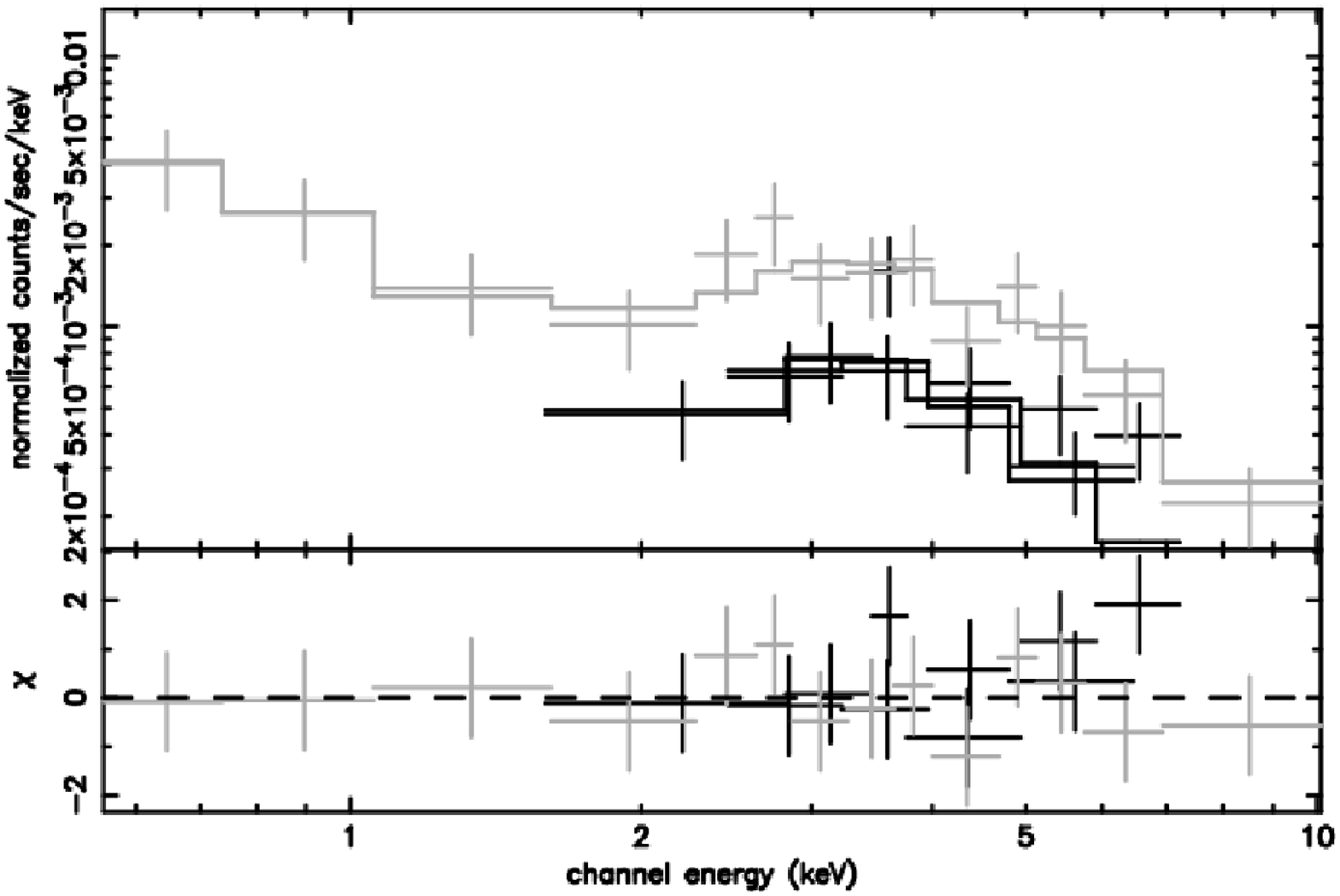}
    \includegraphics[scale=0.35,angle=-90,keepaspectratio]{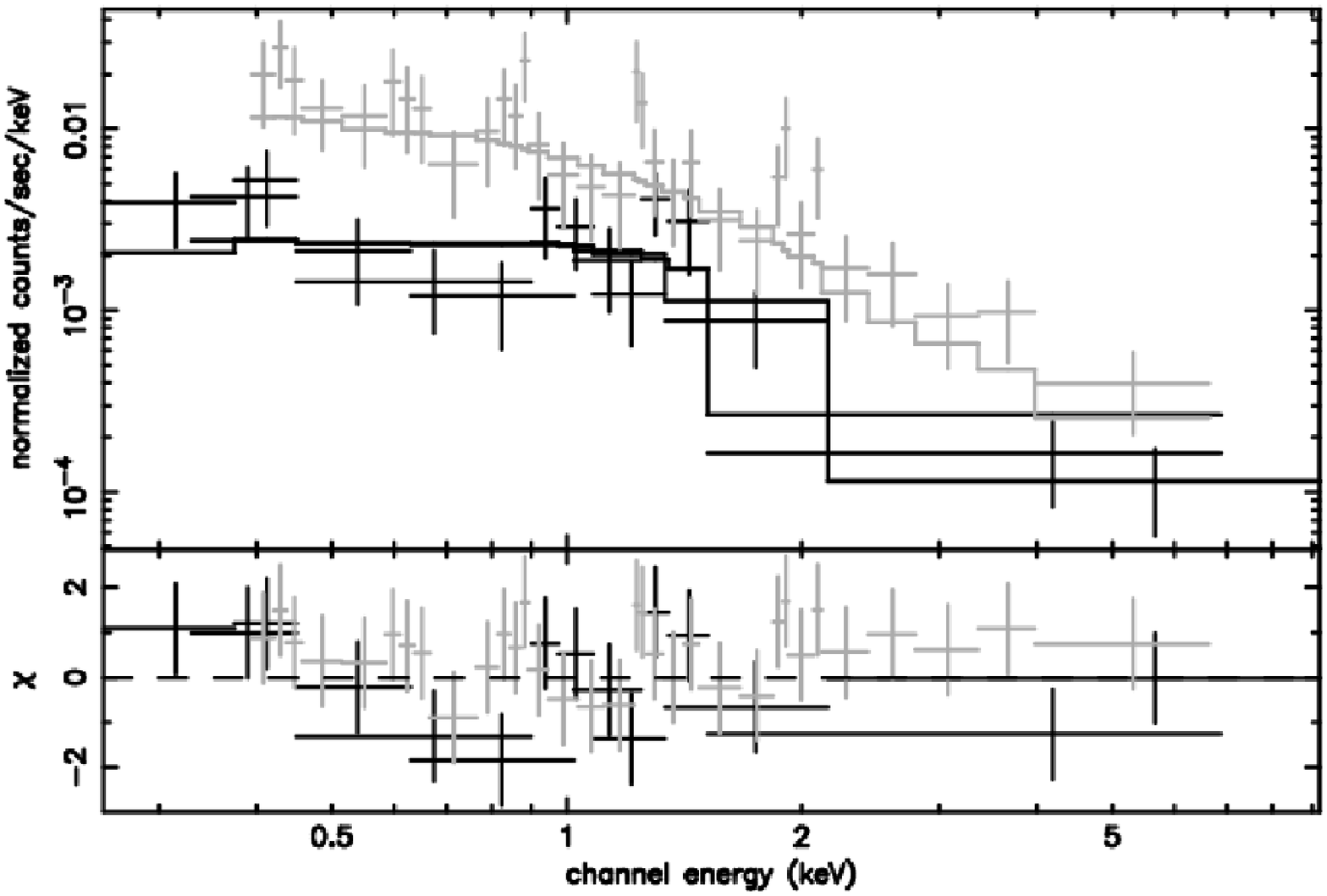}
    \includegraphics[scale=0.35,angle=-90,keepaspectratio]{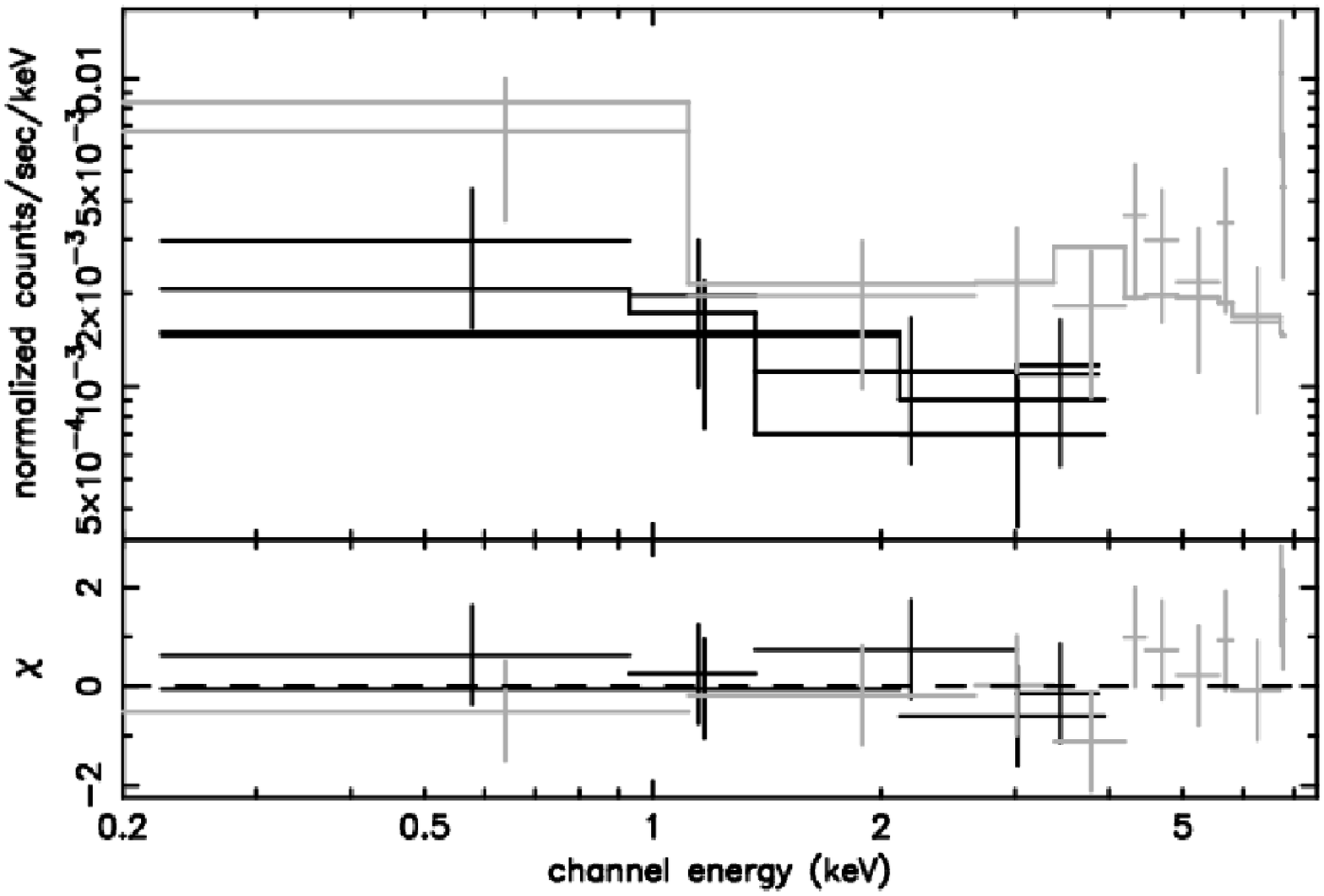}
%\vspace{-0.5cm}
\caption{\small {\bf Top}: Data and  model for the core region of 3C\,292. The model is an  absorbed power law. {\bf Center}: Data and model for the radio lobe region. The model displayed here consists of a power law absorbed by Galactic absorption. {\bf Bottom}: Data and model for the extended emission region, after masking the radio lobe region with two sectors. The model here is a MEKAL of temperature 2.2 keV plus an absorbed power law. MOS data points are in black, pn in grey.}% caption for the whole figure
\label{fig:fig4} 
%\end{minipage}
  \end{center}
\end{figure}

We extracted the spectrum from the core, the lobes and large scale emission (see Fig. \ref{fig:fig4}).
The core spectrum is well fitted with a two power law  model consisting of soft, $\Gamma= 2.6^{+1.5}_{-1.1}$,  and  hard $\Gamma= 2.2^{+0.3}_{-0.2}$, heavily absorbed ($N_H = 2.9\times 10^{23}$ cm $^{-2}$) power laws, which we  relate to the active nucleus and the radio core of the galaxy.

The lobe spectrum gives a best-fit power law model of $\Gamma= 1.88\pm0.26$, from which we derive a flux density at 1 keV of 4.1 nJy. 

By excluding the extended  X-ray emission associated with the lobes and modelling the contribution from the wings of the PSF, we determined a best-fit temperature for the external environment to be k$T = 2.2$ keV.  

We interpret the X-ray emission aligned with the radio lobes  as arising from IC scattering of the CMB, as also supported by the good agreement between the measured X-ray flux and the prediction from  radio data if the source is in equipartition. However some doubts arise about interpreting the structures as  pure non-thermal emission since a thermal model of k$T$ = 5.2 keV is also a good fit. In this case heating of the external medium may be taking place at the location of the lobes. The X-ray emission from the core points to a hidden quasar which displays also a soft component. The large-scale thermal emission indicates the presence of a poor cluster with k$T$=2.2 keV. The external pressure at the radio lobes is $P = 2.5\times10^{-13}$ Pa, which is somewhat low, but consistent with the source being in equilibrium with the external medium.

\subsection{3C 322}
\begin{figure}[!]
  \begin{center}
    \includegraphics[width=\columnwidth, angle=0]{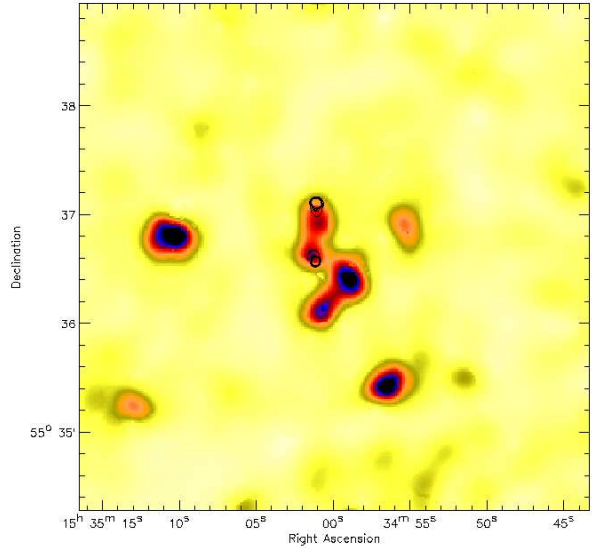}
% Other \includegraphics options include: height, width, angle, scale
    \caption{\small XMM/EPIC adaptive smoothed image of 3C\,322. Contours are from the 1.4 GHz image. }
    \label{fig:fig5}
  \end{center}
\end{figure}

The smoothed image (Fig. \ref{fig:fig5}) is a 10 ks observation, with very high background. We detect only 50 net counts from the source. To the north, the radio emission is more extended than the X-ray. Under the hypothesis that all the photons correspond to the emission from a cluster-like environment we estimate an upper limit for the expected luminosity of any thermal, extended  emission of $L_{\rm X}= 5\times 10^{44}$ erg s$^{-1}$ (assuming k$T$= 4 keV to match the $L_{\rm X}-T$ relation of Vikhlinin et al. 2002)

We obtain rough spectral informations from the 50 counts. A power law of photon index   $\Gamma$ = 1.6 gives a good fit (Cash statistics were used). Adopting this model - thus ignoring any thermal contribution-,  the measured flux density at 1 keV is 1.4 nJy, in very good agreement with the predictions for SSC and IC scattering of the CMB derived from radio data, assuming equipartition between the electrons and magnetic field.

\section{Summary and Conclusions}
We have undertaken an analysis of the X-ray emission from 3 high redshift radio galaxies. Core emission from an AGN is detected for two of them. X-ray emission associated with the radio lobes is spectrally found  for the three sources.
For 3C 184 and 3C 292 we also found cluster like emission of k$T$ =3.6 keV and 2.2 keV, respectively. Neither of them corresponds to a particularly rich cluster, in contrast to some lower-redshift radio galaxies such as Cyg A (Smith et al. 2002) and Hydra A (Nulsen et al. 2002). Adopting our best estimates for the lobe IC emission, the lobes in all three sources are found to be consistent with equipartition within the uncertainties. 
The results are described in more detail in Belsole et al. (2004).

\section*{Acknowledgments}
We thank the organisers of the conference for allowing us to present this work and for the successful organisation of the meeting. E.B. acknowledges support from PPARC. MJH thanks the Royal Society for a fellowship.

\end{document}